\documentclass[%
 reprint,
superscriptaddress,
linenumbers,
 amsmath,amssymb,
 aps,
pra,
]{revtex4-2}
\usepackage[separate-uncertainty=true]{siunitx}
\usepackage{graphicx}
\usepackage{dcolumn}
\usepackage{bm}
\usepackage{hyperref}
\hypersetup{
    colorlinks={true},
    urlcolor={black}, 
    linkcolor={black},
    citecolor={black},
}

\usepackage{siunitx}
\DeclareSIUnit{\dBm}{dBm}
\DeclareSIUnit{\dB}{dB}

\newcommand{\rp}{\mathrm{p}}

\usepackage[dvipsnames]{xcolor}
\newcommand\navyblue[1]{\textcolor{NavyBlue}{#1}}
\usepackage[normalem]{ulem} 
\definecolor{crimson}{rgb}{0.86, 0.08, 0.24}
\newcommand\removed[1]{\sout{\textcolor{crimson}{#1}}}

\iftrue
\renewcommand{\navyblue}[1]{#1}
\renewcommand{\removed}[1]{}
\fi

\begin{document}

\nolinenumbers

\title{Multiplexed readout of \navyblue{Superconductor--Normal-Conductor--Superconductor} bolometers}

\newcommand{\affqcd}{QCD Labs, QTF Centre of Excellence, Department of Applied Physics, Aalto University, FI-00076 Aalto, Finland}
\newcommand{\affvtt}{QTF Centre of Excellence, VTT Technical Research Centre of Finland Ltd., 02044 VTT, Finland}
\newcommand{\affiqm}{IQM Quantum Computers, Espoo 02150, Finland}

\author{Priyank Singh}
\email{priyank.singh@aalto.fi}
\affiliation{\affqcd}

\author{András Gunyhó}\affiliation{\affqcd}
\author{Heikki Suominen}\affiliation{\affqcd}
\author{Giacomo Catto}\affiliation{\affqcd}
\author{Florian Blanchet}\affiliation{\affqcd}
\author{Qi-Ming Chen}\affiliation{\affqcd}
\author{Arman Alizadeh}\affiliation{\affqcd}
\author{Aarne Keränen}\affiliation{\affqcd}
\author{Jian Ma}\affiliation{\affqcd}
\author{Timm Mörstedt}\affiliation{\affqcd}

\author{Wei Liu}\affiliation{\affiqm}
\author{Mikko Möttönen}\affiliation{\affqcd}\affiliation{\affvtt}

\date{\today}

\begin{abstract}

Recently, ultrasensitive calorimeters have been proposed as a resource-efficient solution for multiplexed qubit readout in superconducting large-scale quantum processors. However, experiments demonstrating frequency multiplexing of these superconductor--normal-conductor--superconductor (SNS) sensors are \navyblue{are lacking in the literature}. 
To this end, we present the design, fabrication, and operation of three SNS sensors with frequency-multiplexed input and probe circuits, all on a single chip. These devices have their probe frequencies in the range \SI{150}{\mega\hertz}--\SI{200}{\mega\hertz}, which is well detuned from the heater frequencies of \SI{4.4}{\giga\hertz}--\SI{7.6}{\giga\hertz} compatible with typical readout frequencies of superconducting qubits.  
Importantly, we show on-demand triggering of both individual and multiple low-noise SNS bolometers with very low cross talk. These experiments pave the way for multiplexed bolometric characterization and calorimetric readout of multiple qubits, a promising step in minimizing related resources such as the number of readout lines and microwave isolators in large-scale superconducting quantum computers.

\end{abstract}

\maketitle

\def\thefootnote{o}\footnotetext{These authors contributed equally to this work}\def\thefootnote{\arabic{footnote}}

\section{Introduction}

Low-noise sensors play a central role in improving contemporary physics experiments to new regimes of observations, leading to new discoveries. Especially in quantum-technology applications~\cite{superconductivitybook}, ultrasensitive detectors are of great importance due to extremely low signal levels~\cite{gunyho2024singleshot,oliverqubit_review2109}. In addition, studies of the cosmic microwave background, neutrino mass, dark matter and low dimensional materials call for highly sensitive detectors~\cite{multiplexedbolo2000,freqmultiplexingSQUID2012,IllariJyvaskyla}. The highest sensitivities are achieved by devices operating at cryogenic temperatures, such as superconducting nanowire single-photon detectors (SNSPDs)~\cite{SNSPDfirst,SNSPD_review_2012,SNSPD_review_2021,SNSPD},  transition-edge sensors (TESs)~\cite{TES_first,TES_after50years,TES_bookchapter}, kinetic-inductance detectors (KIDs)~\cite{KID_first,KID_review}, and more recently emerged normal-conductor--insulator--superconductor (NIS) detectors~\cite{JukkaNIS,NIS_asbolometer} and superconductor--normal conductor--superconductor (SNS) sensors, where the normal-conductor part can be either metallic~\cite{Kokkoniemi2019nanobolometer} or a two-dimensional conductor such as graphene~\cite{natureBolometerOperating}.

Advanced applications in quantum technology, as well as multipixel detector arrays in particle physics and cosmology require integrating a large number of detectors in a single experimental setup.
Historically, frequency-domain multiplexing has been a key solution to this challenge.
This technique is well-established for various different types of sensors, but it is at its infancy for ultrasensitive SNS detectors. 

As a specific use case for ultrasensitive detectors, we consider the current major challenge of building useful large-scale superconducting quantum computers, where qubit readout is routinely frequency-multiplexed~\cite{multiplexqubitorigin,maultiplexedqubitwallroff} but scaling is limited by bulky and costly microwave isolators and amplifiers. 
Interestingly, ultrasensitive SNS sensors have been considered~\cite{joonasprl,Kokkoniemi2019nanobolometer} as promising candidates for microwave detection in the framework of superconducting qubits and shown to meet the noise and speed characteristics for typical readout of superconducting qubits~\cite{natureBolometerOperating}. Very recently, an SNS calorimeter was shown to provide reasonable fidelity in single-shot readout of a superconducting transmon qubit~\cite{gunyho2024singleshot}, providing great motivation for demonstrating that these calorimeters can be reliably multiplexed. 

In addition to qubit readout, SNS sensors can be used as a characterization tool for microwaves propagating at millikelvin temperatures \cite{DC_bolometer}. Furthermore, recently we have used an SNS bolometer to measure microwave photon correlations~\cite{keranen2024measurementmicrowavephotoncorrelations}. Multiplexing many bolometers seems tempting also in this scenario, due to the increasing need to quickly characterize cryogenic microwave environments and components for quantum technology \cite{RL_bluefors_bolo}.

The operating principle of SNS sensors is based on the heating of an absorber element by incoming photons, which causes a change in the impedance of SNS junctions.
The input for the absorbed photons is typically engineered to be impedance matched to \SI{50}{\ohm}, and hence the bandwidth of these sensors is naturally much broader than the linewidth of the photons generated by the readout resonators of superconducting qubits.
Thi allows one to channel the readout signals of multiple qubits into a single SNS sensor, but distinguishing the signals corresponding to different qubits is inconvenient since the SNS sensor only measures power, destroying the information about the frequency and phase of the absorbed photons.
To overcome these issues, it was suggested in Ref.~\cite{gunyho2024singleshot} that one SNS sensor is used per qubit such that both the input and probe signals of the sensors are frequency multiplexed, thus requiring only two microwave readout lines for a group of multiplexed readout channels.
However, a detailed study of this scheme, including the characterization of crosstalk both on the input and the probe sides of the sensors, as well as demonstrating simultaneous multiplexing of several sensors has not been reported to date.

In this work, we design, fabricate, and operate a sample with multiplexed excitation and readout of three SNS bolometers on a single chip. The probe frequency of these bolometers is in the range \SI{150}{\mega\hertz} -- \SI{200}{\mega\hertz}, which can be changed by redesigning the capacitors and inductors of the probe circuit and by changing the length of the SNS proximity Josephson junction. In order to individually actuate the bolometers, we have placed on-chip filters at the absorber inputs of each bolometer. Since the principal function of these filters is to isolate the bolometers from each other in the range \SI{4.4}{\giga\hertz} -- \SI{7.6}{\giga\hertz}, the bandwidth of the filters is of great importance. For a bandwidth in the 100-MHz range, we observe less than $\SI{-11.8}{\dB}$ crosstalk between the heater tones corresponding to the different input filters and successfully measure individual and multiple real-time simultaneous excitations of the different bolometers using a single frequency-multiplexed probe tone. This work stands as an important step towards frequency-multiplexed calorimetric readout of superconducting qubits.

\section{Device and operation} 

\begin{figure}[h!]
\centering
\includegraphics[width=0.95\columnwidth]{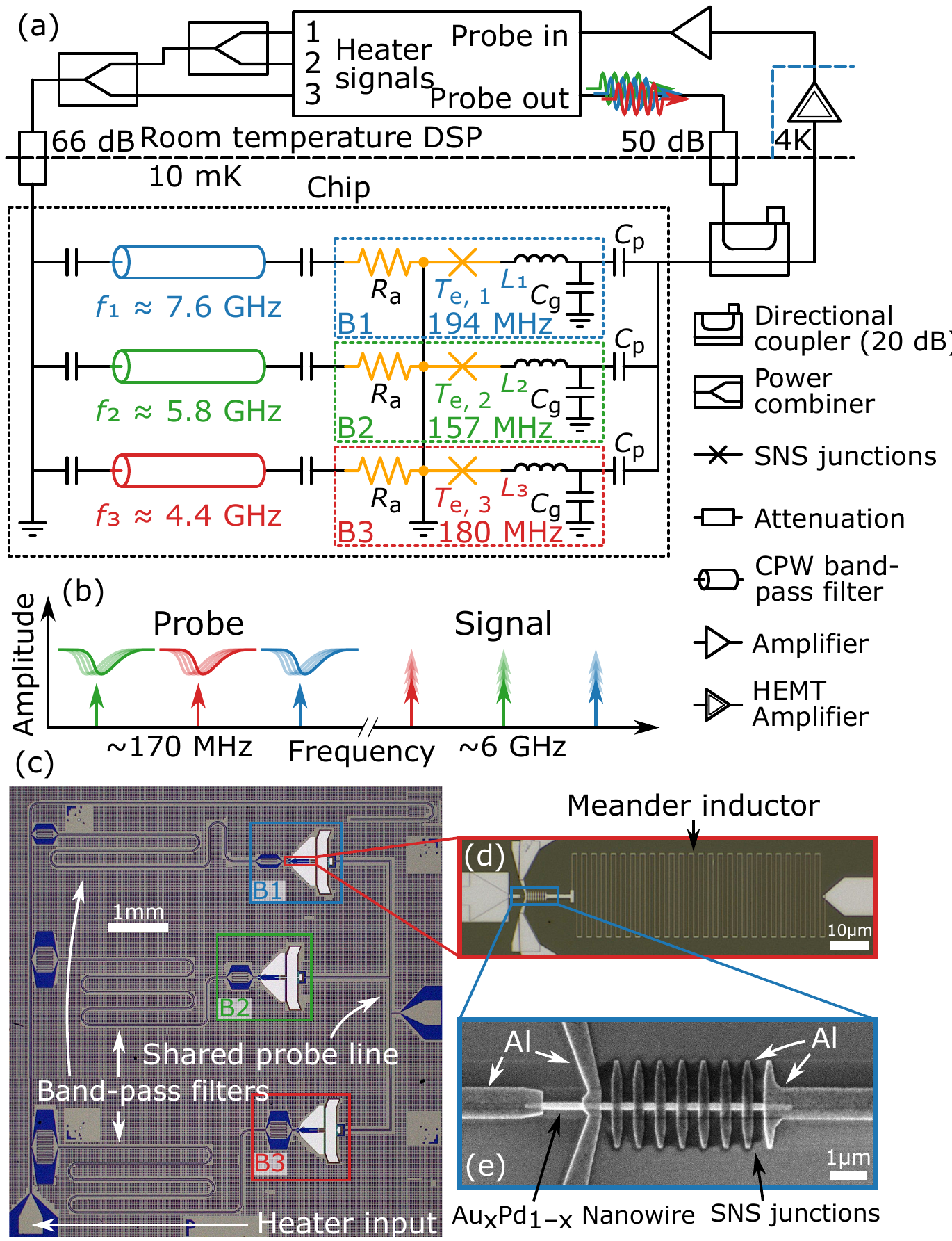}
\caption{%
  \label{fig:threebolometer}
  Experimental setup.
  (a) Simplified diagram of the measurement setup, along with the equivalent circuit diagrams of the bolometers.
  The three bolometers are probed via a shared probe line and the probe signal is amplified by a high-electron-mobility-transistor (HEMT) amplifier at the 4-K temperature stage and by a low-noise amplifier at room temperature.
  The heater signals are combined using power combiners and sent to the on-chip band-pass filters via a shared heater line.
  A digital signal processing (DSP) module handles all signal generation and data acquisition (see text).
  The circuit elements corresponding to each bolometer are highlighted with colored dotted lines.
  The sections highlighted in orange correspond to the nanowires \navyblue{where each cross symbol represent a series of eight superconductor--normal-conductor--superconductor junctions}.
  The parameters that are not nominally identical are colored with the colors corresponding to the different bolometers.
  (b) Illustration of the frequencies involved in the experiment.
  The signal tones are generated near \SI{6}{\giga\hertz}, and changing their power causes a resonance shift near the probe frequencies, which are around \SI{170}{\mega\hertz}.
  (c) Optical micrograph of the chip, with the three bolometers and other components indicated.
  (d) Close-up of bolometer B$_1$, showing the meandering inductor structure.
  (e) Scanning-electron-microscope image of the AuPd nanowire of a nominally identical device to that measured (adapted from Ref.~\cite{gunyho2024singleshot}).
}
\end{figure}

Figure~\ref{fig:threebolometer} shows the experimental setup and the chip with the three bolometers used in this work.
The design of the bolometers is similar to those presented in Refs.~\cite{Kokkoniemi2019nanobolometer, gunyho2024singleshot, joonasprl}.

The \navyblue{approximately \SI{30}{nm} thick and \SI{150}{nm} wide }$\text{Au}_{x}\text{Pd}_{1-x}$ ($x \approx 0.33$) nanowire shown in Fig.~\ref{fig:threebolometer}(e) is used to form both an absorber to absorb the incoming heating power, as well as a thermometer to convert the corresponding temperature change into an electrical signal.
The absorber is formed by a nanowire segment with \navyblue{a length of }\SI{1100}{nm}, chosen such that its impedance is closely matched to \SI{50}{\ohm}~\cite{Kokkoniemi2019nanobolometer}. At the input of the bolometer, the absorber is capacitively coupled to the input line filter and at the other end, the absorber is shorted to ground using a superconducting Al lead. 
The thermometer, on the other hand, consists of a series of superconducting Al islands placed on top of the nanowire, which effectively form superconductor--normal-conductor--superconductor junctions due to the superconductor proximity effect \cite{HolmMeissner1932proximityeffect,meissnerphysrev_proximity,RevModPhys.77.109_proximity}.
The impedance of this weakly superconducting junction chain depends on the electron temperature of the nanowire. The chain is embedded in an $LC$ tank circuit formed by the junctions, a shunt capacitor $C_{\mathrm{g}}$ \navyblue{(134 pF)}, and a series meander inductor $L_{j}$, $j=1,\, 2,\, 3$ [Fig.~\ref{fig:threebolometer}(d)], where the index $j$ refers to the specific bolometer B$_1$, B$_2$, and B$_3$, respectively.
The resonance frequency of the $LC$ circuit thus depends on the electron temperature of the nanowire and therefore on the absorbed radiation power. Changes in the resonance frequency are probed in a microwave reflection configuration.

To multiplex the readout of the bolometers, we place three individual devices described above on the same chip with a shared probe line as shown in Fig.~\ref{fig:threebolometer}(c).
The inductance of each inductor $L_{j}$ is tuned in fabrication by adjusting the length in order to separate the resonance frequencies of the three tank circuits from each other.
On our chip, we also share the heater input line between the bolometers.
Therefore, in order to individually address the inputs, we couple each absorber to the heater input line through half-wavelength coplanar-waveguide (CPW) filters~\cite{goppl_CPW}.

As shown in Fig.~\ref{fig:threebolometer}(a), from the room-temperature electronics, we send a probe signal through attenuators and a directional coupler.
The reflected signal from the device travels to a HEMT amplifier, the output of which is read by a digital signal processing module. 
 In principle, the probe output signal is a reflected signal from the bolometer.  However, from the point of view of the room-temperature electronics, the signal is measured in transmission due to the directional coupler. Hence, we refer to it as the transmitted signal.
In the initial characterization measurements, the room-temperature experimental setup is essentially identical to that in Ref.~\cite{joonasprl}, using analog heterodyne up- and downconversion.
However, for the multiplexed measurements discussed below, we use a fully digitally synthesized microwave source and analyzer module~\cite{prestopaper}.

\section{Sample characterization}
\label{sec:characterization}

    \subsection{Transmission measurement of bolometers} 

Figure~\ref{fig:characterization}(a)--(c) shows the magnitude of the signal reflected from the probe tank circuits of each of the three bolometers. The resonances at a nominal probe power of $-144$~dBm at the device input are observed to be at 193.7~MHz (B$_1$), 156.7~MHz (B$_2$), and 179.3~MHz (B$_3$). 
Note that the ascending order of the tank circuit resonance frequencies do not follow the order of the associated filter frequencies.

The resonance frequencies of the bolometers decrease with increasing probe power, which we attribute to electro-thermal feedback~\cite{joonasprl}.
With a probe power above approximately \SI{-125}{\dB}, the bolometers enter a nonlinear regime, where they are not stable, and hence we limit our operation point to be in the linear regime of weak probe power.
For each bolometer, we optimize the operation point such that the feedback does not cause nonlinear effects, since they cause instability.
Namely, we fix the probe power at \SI{-144}{\dBm} for all the bolometers unless otherwise explicitly stated. Furthermore, we operate the bolometers B$_{1,2,3}$ at 194.0~MHz, 156.6~MHz, and 180.2~MHz, respectively.
For the sake of brevity of presentation, we round these numbers to single megahertz for rest of the article.
\navyblue{The linewidths of these bolometers are \SI{0.61(0.03)}{\mega\hertz}, \SI{0.31(0.02)}{\mega\hertz} and \SI{0.14(0.01)}{\mega\hertz}, respectively.}

Even though the bolometers share a common readout line, the operation point for each of them remains independent due to the large separation between the resonances compared with their linewidths.

 \subsection{Filter characterization}

\begin{figure}[h]
\includegraphics[width=\columnwidth]{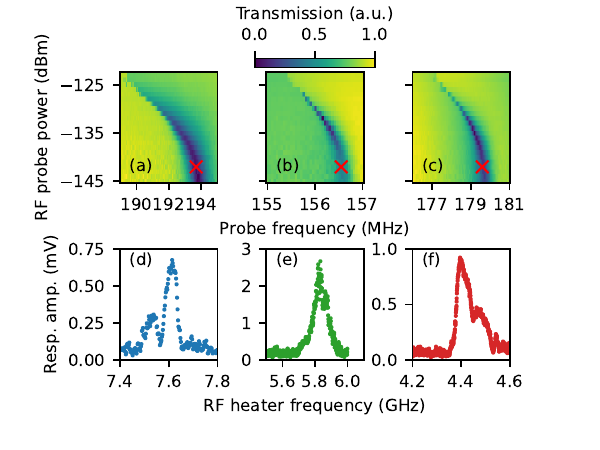}
\caption{
    \label{fig:characterization} Sample characterization.
    (a)--(c) Measured transmission amplitude as a function of the probe frequency $f_{\rp}$ and the probe power $P_{\rp}$ near the resonance frequency of the bolometer (a) B$_1$ \SI{193.7}{\mega\hertz}, (b) B$_2$ \SI{156.7}{\mega\hertz}, and (c) B$_3$ \SI{179.3}{\mega\hertz} tank circuits.
     The amplitude in each panel is normalized to the range $[0, \, 1]$ for clarity.
    At a nominal probe power of approximately $\SI{-140}{\dBm}$, the frequency change with power is significant. 
    (d)--(f) Measured change in the transmission amplitude of the probe signal of bolometer (d) B$_1$, (e) B$_2$, and (f) B$_3$ as a function of the heater frequency in the vicinity of the pass band of the input filter of the corresponding bolometer. The bolometer shows maximum change in the transmission amplitude at the resonance frequencies of the filters.
}
\end{figure}

In order to characterize the resonance frequencies of the filters, we measure the transmission at specific operation points, determined by the probe frequency and probe power of each bolometer.
We send microwave pulses at different frequencies through the heater input port of the chip and monitor the resulting change in the probe signal at each of the bolometer operation points.
We refer to this change in the transmission coefficient of the probe tone as the response of the detector, or detector response.
Figure~\ref{fig:characterization}(d)--(f) shows the result of these measurements around the passband of each of the filters.
For each filter, we study the response of the corresponding bolometer. Note that we may also observe a dip, i.e., negative response, instead of a rise, i.e., positive response, depending on the operation point of the bolometer.
The response of each bolometer depends on the strength of the heater signal.

In Fig.~\ref{fig:characterization}(d)--(f), we find the filter center frequencies $f_1 = \SI{7.6}{\giga\hertz}$, $f_2 = \SI{5.8}{\giga\hertz}$, and $f_3 = \SI{4.4}{\giga\hertz}$, corresponding to the bolometers B$_1$, B$_2$, and B$_3$, respectively. \navyblue{The strongest peaks of the filters have linewidths of approximately \SI{100}{\mega\hertz}.}
The filter frequencies are engineered for the range of $4$--$\SI{8}{\giga\hertz}$, which is typical in circuit quantum electrodynamics~\cite{blais2021circuitQED,gunyho2024singleshot} experiments.
Note that the filter transmission curves exhibit significant deviations from their expected Lorentzian shape, which can be attributed to impedance mismatches both between the 50~$\Omega$ input of the chip and the on-chip filter lines, as well as between the filters and the AuPd absorber.
\navyblue{To understand the origin of such behavior, we carried out numerical microwave simulations of the design. We discuss the effect of the impedance mismatch on the shape and cross talk of the filters in the SI (Fig.~1 of SI).} In future work, we aim to improve these filters, for example, by employing lumped-element filters instead of CPW-based filters~\cite{lumpedelementqcd}. 
Despite the distortion in the filter lineshapes, they are sufficient to isolate the heater signals for convenient actuation of the individual bolometers as we show below. %

\section{Frequency multiplexing}
    
    \subsection{Individual trigger measurements}

Before carrying out multiplexed readout, we study each of the bolometers separately to ensure that operating one bolometer does not impact the operation of the other bolometers.
To this end, we first apply a heater pulse to each bolometer individually, while monitoring all the bolometers one by one.

Note that in this and the following sections, we present results in increasing order of bolometer tank circuit resonance frequency.

\begin{figure}
\includegraphics[width=\columnwidth]{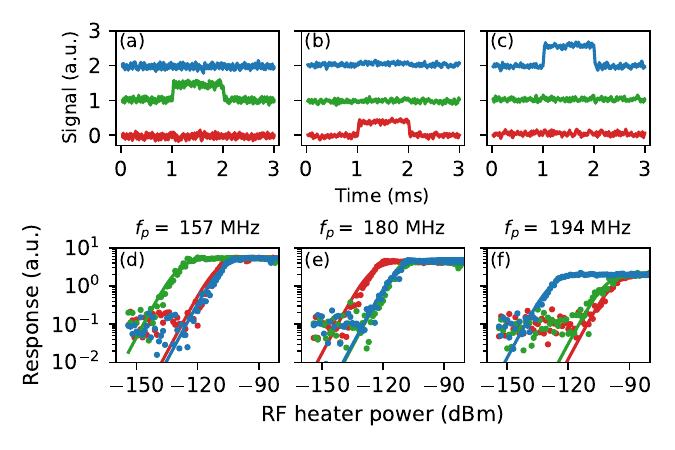}
\caption{
    \label{fig:singleTrigger}
    Power dependent time domain measurements by triggering individual bolometers. (a)--(c) Time traces of the transmission amplitudes of the probe tone at (a) \SI{157}{\mega\hertz}, (b) \SI{180}{\mega\hertz}, and (c) \SI{194}{\mega\hertz} for a heater pulse of \SI{5.8}{\giga\hertz} (green), \SI{4.4}{\giga\hertz} (red), and \SI{7.6}{\giga\hertz} (blue) at a constant heater power of \SI{-135}{\dBm}. The curves are vertically offset from each other for clarity.
    (d)--(f) Bolometer response amplitude at (a) \SI{157}{\mega\hertz}, (b) \SI{180}{\mega\hertz}, and (c) \SI{194}{\mega\hertz} probe frequency from experiments similar to those of (a)--(c) as a function of the heater power applied at the center frequency of the filter corresponding to the bolometer being probed. 
    The heater power dependence shows that the filters are separating the heater signal reaching to the respective bolometers. 
    %
}
\end{figure}

To implement these individual-trigger measurements, we apply a continuous probe tone for a chosen bolometer and monitor the transmission coefficient from the sample. A heater pulse shorter than the measurement sequence is sent through the heater port with a time delay. We bandpass filter the probe signal at room temperature in Ref.~\cite{joonasprl}. The overall measurement of the transmission coefficient captures the event of heating in time domain. Figure \ref{fig:singleTrigger}(a)--(c) shows the time-domain transmission of each of the three bolometers with \SI{1}{\milli\second} heater pulses applied at the center frequencies of each of the filters, with a relatively low heating power of $\SI{-135}{\dBm}$. We collect $2^{14}$ averages of each time trace for sufficient signal-to-noise ratio (SNR) to capture the crosstalk of the heater signals between the bolometers. Even with such large averaging, we observe that the heater pulse is only detected by the desired bolometer, indicating that the heater signals are isolated at the selected power. We used similar measurements to estimate the time constants of these bolometers which are found to be in the range of 4--\SI{13}{\micro\second} (see Supplementary Information).

Although at a low power level the cross talk between the heater signals is not visible, it is expected that the the heater tone at the resonance frequency of one filter leaks through the other filters, the effect of which is expected to become measurable at high heater powers.
In order to quantify this leakage, we monitor the heater power dependent detector response for each of the bolometers and for the different filter frequencies, as shown in Fig.~\ref{fig:singleTrigger} (d)--(f).
We fit these data to a phenomenological model of the power dependence of the output signal introduced in Ref.~\cite{gunyho2024singleshot}.
From these fits, we extract the $P_{\SI{1}{\dB}}$ compression point, which describes the saturation of the detector response with respect to the power applied through each filter.
The difference between the compression points can be used to determine the amount of crosstalk between the filters.
The parameters from the fits are shown in Table~\ref{p1db}.
Overall, we find that the crosstalk is between \SI{-11.8}{\dB} and \SI{-17.9}{\dB}. This means that we have a range of roughly \SI{12}{\dB} for safe operation, limited by $f_1$. 

\begin{table}[]
    \centering
        \caption{ \navyblue{One-decibel compression point of the probe signal of each bolometer with respect to the heater tone which is applied at the indicated frequencies of the input filters in units of dBm. The shown 1$\sigma$ uncertainties are derived from the covariance of the fit parameters.}}
    \label{p1db}
    \begin{tabular}{|c|c|c|c|}
    \hline
Bolometer / Filter  & \SI{4.4}{\giga\hertz} &\SI{5.8}{\giga\hertz} & \SI{7.6}{\giga\hertz} \\  \hline
 B$_1$      & $-106.3  \pm 0.3$        & $-101.9 \pm 0.4$       & $-128.2 \pm 0.4$     \\   \hline
B$_2$     & $-114.3 \pm 0.4$        & $-135.5 \pm 0.4$       & $-116.4 \pm 0.7$    \\    \hline
 B$_3$      & $-132.0 \pm 0.3$          & $-120.0 \pm 0.4$         & $-120.0 \pm 0.3$    \\      \hline

    \end{tabular}
\end{table}

Importantly, the level of crosstalk is determined by the filter quality. 
In the following, we use heater powers around $\SI{-135}{\dBm}$, which minimizes the effect of the crosstalk while maintaining a good SNR for all bolometers.

    \subsection{Simultaneous frequency domain multiplexing}

For a simultaneous frequency-multiplexed readout of all the bolometers, we create three pulses with independently specified amplitudes at the operating frequencies of the bolometers and digitally combine them before synthesizing. The heater pulses are synthesized individually and combined  at room temperature using physical power combiners as shown in Fig.~\ref{fig:threebolometer}(a). 
We label each experiment with a three-digit binary number 000, 001,.., 111, where \navyblue{the bits are ordered from left to right in order of increasing bolometer probe frequency and the digits 0 or 1 represent the heater off and on states of the corresponding bolometer, respectively}.

For example, an experiment where a heater pulse is sent at the heater frequency of the \SI{194}{\mega\hertz} bolometer while no pulse is sent at the two other heater frequencies corresponds to the binary number 001.

\begin{figure}
\includegraphics[width=\columnwidth]{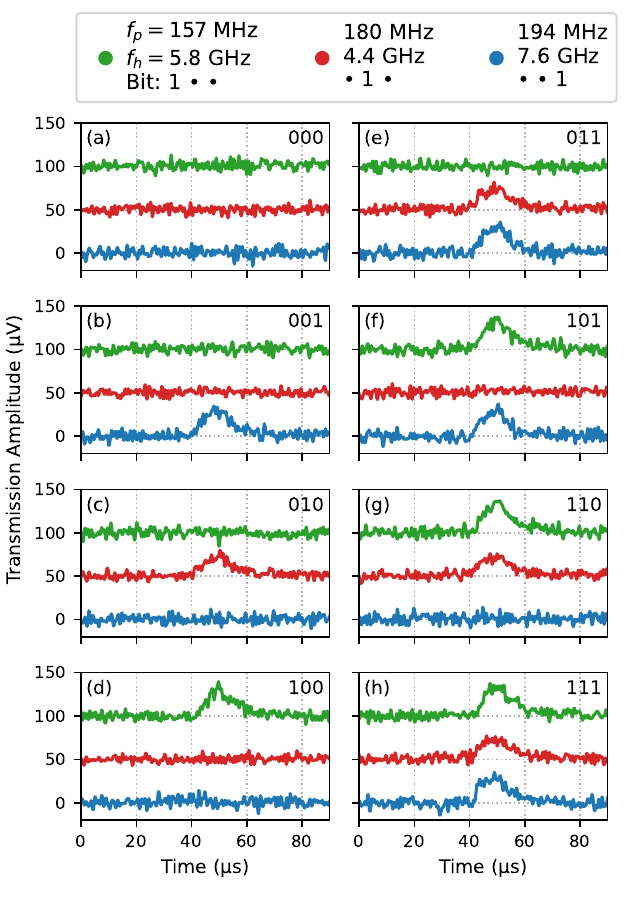}
\caption{\label{fig:FT_demux} Frequency-multiplexed simultaneous readout of the bolometers.
        \navyblue{
    (a) Transmission amplitude of the multiplexed probe signal demodulated at probe frequencies of~\SI{157}{\mega\hertz} (green), \SI{180}{\mega\hertz} (red), and \SI{194}{\mega\hertz} (blue),  with all heaters turned off.
    For clarity, the \SI{180}{\mega\hertz} and \SI{157}{\mega\hertz} data are offset by 50 and \SI{100}{\micro\volt}, respectively, in all panels.
    The matching heater frequencies and the bit they correspond to are shown in the legend.
    The bit pattern shown in each panel is indicated in the top right corner.
    (b)--(d) Transmission amplitudes of all three bolometers for individual heaters turned on.
    (e)--(g) Transmission amplitudes of all three bolometers for two heaters turned on simultaneously.
    (h) Transmission amplitude for all three heaters turned on.
    }
   }
\end{figure}
\navyblue{For these multiplexing experiments, }we reduce our measurement window to a duration of \SI{100}{\micro\second} and the duration of the heater pulses down to \SI{10}{\micro\second}, applied \SI{40}{\micro\second} after the beginning of the measurement 
\navyblue{window. We omit the data of first \SI{10}{\micro\second} of the measurement window for clarity. 
This resembles practical applications of pulse or particle detection by calorimetric operation, although, for example, in qubit readout, the pulse durations are on the order of \SI{100}{\ns}, or in cosmic ray detection, energy is deposited effectively instantaneously in a single event.
We choose to not reduce the pulse duration further below the 4--\SI{13}{\us} thermal time constants of the bolometers in order to maintain a reasonable SNR.}

The transmitted signal is digitized at a sampling rate of \SI{6}{GSa/s}. We digitally filter the acquired data using a \SI{1}{\mega\hertz} brick-wall band-pass filter at each of the operation frequencies of the bolometers to recover the individual signals.
We record the time domain signal for for each possible combination of heater on and off states with $10^4$ averages to ensure a reasonable SNR and to monitor any leakage \navyblue{from} the heater signal to the non-addressed bolometers.

\removed{The result of this process is shown in Fig.~\ref{fig:FT_demux}(a)--(h).} 
\navyblue{
Figure \ref{fig:FT_demux} shows the time-domain transmission data for different possible combinations of actuating the bolometers.
Figure~\ref{fig:FT_demux}(a) shows the state of different bolometers while none of the heater pulses are sent.
Figures~(b)--(d) present the data for individual actuation of the bolometers while keeping the other heaters off.
Figures~\ref{fig:FT_demux}(e)--(g) present the case of simultaneously actuating two bolometers and Figure~\ref{fig:FT_demux}(h) shows the corresponding data for all three heater pulses applied.
A clear response is visible in all the bolometers at \SI{40}{\micro\second} after the beginning of the measurement sequence, when the respective heaters are turned on.
We demonstrate that the transmission amplitudes for all the bolometers remain essentially unresponsive unless a heater signal at the corresponding heater frequency actuates the bolometer, regardless of the state of other heater signals.
}%
There are two sources of possible crosstalk between the bolometers: (i) the probe signal of a non-addressed bolometer and (ii) the heater signal of a non-addressed bolometer. Let us analyze each of these cases below.

For case (i), as we observe in Fig.~\ref{fig:FT_demux}, the change in transmission amplitude of a certain probe signal is insignificantly affected by other probe signals. We attribute the low crosstalk of the probe signals to the large separation of the resonance frequencies of the probe tank circuits compared with their linewidths and to the low-power operation point of the bolometers evading the crosstalk owing to non-linearity. 
Furthermore, the bolometers are probed at a frequency of around \SI{150}{\mega\hertz}, corresponding to a wavelength of more than a meter, which places our device well within the lumped-element domain. 

\begin{table}
    \centering
    \caption{
    \navyblue{
    Mean of the time-domain transmission signal averaged over a time interval of 47--52 \si{\micro\second} after the beginning of the measurement divided by the standard deviation before the pulse (0--30 \si{\micro\second}) for each of the bolometers. The top four rows correspond to the leakage of signal from filters corresponding to the other bolometers and the bottom row shows the value with only the corresponding heater on, i.e., the signal-to-noise ratio.
    }
    }
    \label{tab:std_dev}
    \begin{tabular}{|c|c|c|c|}
    \hline
        $f_p$ & 157~MHz & 180~MHz & 194~MHz \\ \hline
        All heaters off & 0.23 & 0.04 & 0.18 \\ \hline
        One heater on & 0.49 & 0.40 & 0.19 \\ \hline
        The other heater on & 0.25 & 0.06 & 0.29 \\ \hline
        Both heaters on & 0.10 & 0.50 & 0.27 \\ \hline
        SNR & 8.38 & 6.99 & 7.47 \\ \hline
    \end{tabular}
    
\end{table}

For case (ii), note that the transmission amplitude around one triggered bolometer does not significantly change due to heater signals targeting other bolometers. These values only change when the respective heaters are triggered individually or \navyblue{in combination} with other heaters\navyblue{, as evident from Fig.~\ref{fig:FT_demux}}.
\navyblue{
To quantify this observation, Table \ref{tab:std_dev} presents the time-domain probe signal of Fig.~\ref{fig:FT_demux} averaged over 47--52 \si{\micro\second} and divided by its standard deviation over time before the heater pulse for each bolometer.
The leakage is lower by more than an order of magnitude than the actual signal.
}

Importantly, these multiplexed bolometers can be triggered by a \removed{short} heater pulse of \removed{only} \SI{10}{\micro\second}, similar to the time scales used, for example, for qubit readout using bolometers~\cite{gunyho2024singleshot}. \navyblue{Note that the calorimetric mode of operating these bolometers typically assumes much shorter heater pulses compared with the thermal time constant. In Ref.~\cite{gunyho2024zeptojoulecalorimetry}, the thermal time constant of the bolometer was roughly \SI{260}{\micro\second}, and hence a much shorter heater pulse of \SI{1}{\micro\second} may be used for making single-shot experiments. In case of qubit readout, the thermal time constant of the used bolometer was ($\tau_b\approx\SI{36}{\micro\second}$)~\cite{gunyho2024singleshot}, much longer than the values reported here, which indicates that much faster readout should be possible with these devices than those of Ref.~\cite{gunyho2024singleshot}, if used in calorimetric mode}.
\navyblue{Note that bolometers based on graphene can outperform the metallic bolometers in terms of thermal speed. For example, typical thermal time constant of a graphene bolometer is around \SI{300}{\nano\second}~\cite{Kokkoniemigraphene2020}, much shorter than the time constants observed above.}

Note that these measurements correspond to a calorimetric operating mode where the heater pulse is shorter or comparable in length to the thermal time constant of the detector since we do not observe a steady-state response for each detector. This suggests that this device may also work as a multiplexed calorimeter although the detailed analysis of the frequency-multiplexed calorimetric mode is left for future research.
\newline

\section{Conclusion and outlook}
In this work, we designed and fabricated on-chip bolometers along with frequency-engineered band-pass input filters and probe tank circuits to selectively actuate and read out the bolometers.
We observed that the crosstalk between the heater signal applied at the filters was roughly between \SI{-12}{\dB} to \SI{-18}{\dB}, and found virtually no interference between the probe signals of the devices at the chosen operation points. 
We have realized the simultaneous multiplexed readout of three on-chip bolometers, detecting microwave pulses in the range of \SI{4}{\giga\hertz} to \SI{8}{\giga\hertz}.

\navyblue{
Our bolometers are probed with frequencies near \SI{150}{\mega\hertz}.
In previous work, similar devices have been demonstrated with probe frequencies above \SI{950}{\mega\hertz}~\cite{russel_aem}, hence up to one~\SI{ }{\giga\hertz} is feasible.
Considering an overall frequency shift of \SI{5}{\mega\hertz}, and a typical linewidth of \SI{0.5}{\mega\hertz},
we estimate that 180 such bolometers can be multiplexed on a single chip in the range of \SI{100}{\mega\hertz} to one~\SI{}{\giga\hertz} without intolerable frequency crowding.%
}
Our results \navyblue{thus} pave the way for large arrays of highly sensitive SNS detectors, which may find applications in large-scale superconducting quantum computing, as well as in fundamental research such as that of radio astronomy.

\begin{acknowledgments}
PS and AG contributed equally to this work. This work was supported by Future Makers, OpenSuperQ, ConceptQ, QTF, and QuTI. PS acknowledges Matti Hokkanen and Leif Grönberg for their assistance in fabrication. Special thanks to Dr. Suman Kundu and Mr. Aashish Sah for discussions. 
\end{acknowledgments}

\bibliography{bibliography}

\end{document}